\documentclass[aps,prb,twocolumn]{revtex4}

\usepackage{amssymb}
\usepackage{graphics}
\usepackage{graphicx}
\usepackage{amsmath}

\begin{document}

\title{On the decrease of intermittency in decaying rotating turbulence}

\author{J. Seiwert}
\author{C. Morize}
\author{F. Moisy}\email{moisy@fast.u-psud.fr}

\affiliation{Univ Paris-Sud 11, Univ Pierre et Marie Curie - Paris 6, CNRS. Lab FAST, B\^at 502, Campus Universitaire, Orsay F-91405, France}

\def\etal{\mbox{\it et al.}}

\date{\today}

\begin{abstract}
The scaling of the longitudinal velocity structure functions, $S_q(r) = \langle \left| \delta u (r) \right|^q \rangle \sim r^{\zeta_q}$, is analyzed up to order $q=8$ in a decaying rotating turbulence experiment from a large Particle Image Velocimetry (PIV) dataset.  The exponent of the second-order structure function, $\zeta_2$, increases throughout the self-similar decay regime, up to the Ekman time scale. The normalized higher-order exponents, $\zeta_q / \zeta_2$, are close to those of the intermittent non-rotating case at small times, but show a marked departure at larger times, on a time scale $\Omega^{-1}$ ($\Omega$ is the rotation rate), although a strictly non-intermittent linear law $\zeta_q / \zeta_2 = q/2$ is not reached.

\end{abstract}

\maketitle

Whether intermittency of isotropic three-dimensional (3D) turbulence is decreased or even suppressed in the presence of system rotation has recently received a marked interest.\cite{Baroud02,Muller07} Here, intermittency refers to the anomalous scaling of the structure functions (SF) of order $q$, $S_q(r) = \langle \left| \delta u (r) \right|^q  \rangle \sim r^{\zeta_q}$, where $\delta u({\bf x},r) = [{\bf u} ({\bf x+r}) - {\bf u} ({\bf x})] \cdot {\bf r} / r$ is the longitudinal velocity increment, ${\bf r}$ an inertial separation normal to the rotation vector ${\bf \Omega}$ and $\langle \cdot \rangle$ denotes spatial and ensemble average. A linear variation of the exponents $\zeta_q$ with the order $q$ is the signature of self-similar (non-intermittent) velocity fluctuations, a situation which is found in the inverse cascade of two-dimensional (2D) turbulence.\cite{Boffetta00} On the other hand, anomalous exponents, $\zeta_q / \zeta_2 \neq q/2$, are the landmark of 3D isotropic turbulence.\cite{Benzi93,She93,Frisch95}  Based on the qualitative ground that  rotating turbulence experiences a partial two-dimensionalization, one may naively expect a reduction or a suppresion of intermittency by comparison with the 3D non-rotating case. More precisely, describing rapidly rotating turbulence in the limit of zero Rossby numbers as a sum of weakly interacting random inertial waves, the vanishing of non-linear effects should lead to a special case of non-intermittent wave turbulence.~\cite{Newell01,Bellet06}

Two papers have recently addressed the issue of the scaling of the SF in rotating turbulence with a stationary forcing. The hot-wire measurements of Baroud et al.\cite{Baroud02} in a turbulent flow generated by radial jets in a rotating tank showed a transition from an intermittent to a non-intermittent behavior, characterized by a $E(k) \sim k^{-2}$ energy spectrum (i.e. $\zeta_2 = 1$) and linear higher-order exponents $\zeta_q  = q/2$. In a Direct Numerical Simulation (DNS) of rotating turbulence with a large-scale isotropic forcing, M\"uller and Thiele\cite{Muller07} have observed reduced intermittency, also characterized with  $\zeta_2 \simeq 1$, but higher-order exponents $\zeta_q$ intermediate between $q/2$ and the values usually found in classical (intermittent) 3D turbulence. Those observations are in qualitative agreement with the increase of $\zeta_q$ reported by Simand\cite{Simand02} from hot-wire measurements in the vicinity of a strong vortex, although no clear separation between a constant background rotation and an otherwise homogeneous turbulence advected by the rotation can be defined in this geometry.  To date, no theoretical description of the scaling of the anisotropic higher order SF in rotating turbulence is available. Note that, in all the above references, only separations ${\bf r}$ normal to the rotation vector ${\bf \Omega}$ are considered, ignoring the complexity originating from the anisotropic character of rotating turbulence.\cite{Bellet06}

In this Letter we report new measurements of the high order SF, carried out by Particle Image Velocimetry (PIV), in a freely decaying rotating turbulence experiment, aiming to compare with the results obtained in forced turbulence.  The experimental setup is the same as in Morize et al.,\cite{Morize05} and is only briefly described here. It consists in a water filled glass tank of square section, of side 35~cm and height $h=44$~cm, rotating at constant angular velocity.  After the fluid is set in solid body rotation, turbulence is generated by towing a co-rotating square grid, of mesh size $M = 3.9$~cm, at a constant velocity $V_g = 0.65$~m~s$^{-1}$ from the bottom to the top of the tank, and is maintained fixed near the top during the decay of turbulence. The horizontal components of the velocity fields in a centered horizontal area of 17 cm $\times$ 14 cm at mid height of the tank are obtained using a corotating PIV system operating at 1~Hz. The velocity fields are defined on a $160 \times 128$ grid, with a spatial resolution of 1~mm and a signal-to-noise ratio of about $2 \times 10^{-2}$.  Although this fails to resolve the dissipative scales (the Kolmogorov scale is approximately 0.2~mm in the first period of the decay), this resolution allows us to resolve the inertial range, typically for $r > 10$~mm.

\begin{table}[b]
\caption{\footnotesize Non-dimensional parameters for the two rotation rates. $[t_0, t_c]$ is the range of approximately self-silimar energy decay.\cite{Morize06}  $Re_g$ and $Ro_g$ are the grid Reynolds and Rossby numbers. $Re_M (t) = u'(t) M / \nu$, $Ro_M(t) = u'(t) / (2 \Omega M)$ and $Ro_\omega(t) = \omega'(t) / 2\Omega$ are the instantaneous Reynolds, macro- and micro- Rossby numbers respectively, based on the horizontal velocity rms $u'(t)$ and vertical vorticity rms $\omega'(t)$.}
\footnotesize
\begin{tabular}{p{34mm}p{24mm}p{24mm}}
\hline \hline
$\Omega$ (rad~s$^{-1}$) & 1.13 & 2.26 \\
$Re_g = V_g M / \nu$ & $2.5 \times 10^4$ & $2.5 \times 10^4$ \\
$Ro_g = V_g / (2 \Omega M)$ & 7.4 & 3.7 \\
$\Omega t_0 /  2 \pi  \dots \Omega t_c /  2 \pi$ & 1.2 \dots 7.4 & 0.6 \dots 10.5 \\
$Re_M (t=t_0 \dots t_c)$ & 1300 \dots 360 & 1400 \dots 380\\
$Ro_M (t=t_0 \dots t_c)$ & 0.38 \dots 0.10 & 0.21 \dots 0.056\\
$Ro_\omega (t=t_0 \dots t_c)$ & 2.1 \dots 0.23 & 1.1 \dots 0.17\\
\hline \hline
\end{tabular}
\label{tab:fp}
\end{table}

Two angular velocities have been used in the present experiments, $\Omega = 1.13$ and 2.26~rad.s$^{-1}$. The corresponding non-dimensional parameters are summarized in Table~\ref{tab:fp}. The grid Reynolds number is $Re_g = V_g M / \nu = 2.5 \times 10^4$ ($\nu$ is the kinematic viscosity), and the grid Rossby numbers $Ro_g = V_g / (2 \Omega M)$ are 7.4 and 3.7, so that the initial state can be considered as a fully developed 3D turbulence weakly affected by the system rotation. A previous investigation\cite{Morize06} showed that, for those rotation rates, the energy decay was approximately self-similar between $t_0 \simeq 40 M / V_g$ and $t_c \simeq 0.10 t_E$, where $t_E = h / (\nu \Omega)^{1/2}$ is the Ekman time, followed by an exponential decay at larger times. The present investigation is restricted to the self-similar range $[t_0, t_c]$. The instantaneous Reynolds, macro- and micro- Rossby numbers, $Re_M$, $Ro_M$ and $Ro_\omega$ respectively, are also given for these two limiting values $t_0$ and $t_c$  in Table~\ref{tab:fp}.

To ensure proper convergence of the statistics, each decay is repeated approximately 600 times, representing 10 hours of run for each rotation rate. It is worth pointing that computing SF from PIV data requires special care, especially when higher order are considered, for which even a small number of spurious vectors may have a large effect. Since those bad vectors may be preferentially found in regions of large velocity or large gradient, finding correct criteria for removing them without introducing biases is a delicate issue. In particular, some of the fields were found to suffer from an inhomogeneous lighting, because the imaged area was partially shadowed when the corner of the tank passed through the laser sheet. Using a criteria based on the $Q$-factor (ratio of primary and secondary correlation peak), 20\% of the fields were aftected by this problem and have been removed. A median filter is then applied to the remaining fields, and it was checked that the SFs computed from the raw and median-filtered data agreed within the error bars $\Delta S_q$ defined below.

\begin{figure}[t]
\centerline{\includegraphics[width=6.5cm]{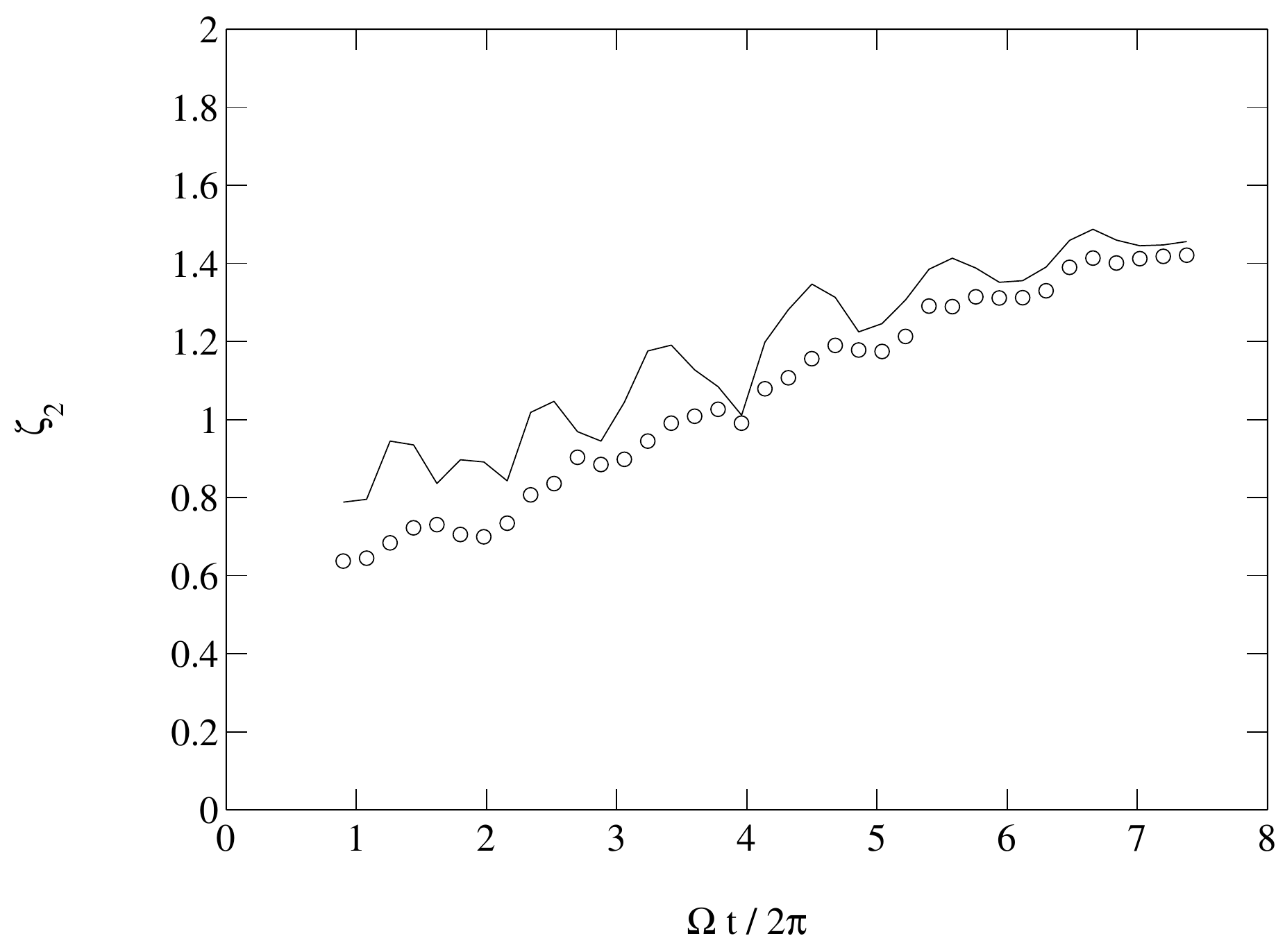}}
\caption{\footnotesize Time evolution of the second order exponent $\zeta_2$ for $\Omega = 1.13$~rad~s$^{-1}$. ---, whole velocity field; $\circ$, turbulent field (ensemble average substracted).}
\label{fig:exp_2_180}
\end{figure}

We first focus on the time evolution of the exponent $\zeta_2$ of the second order SF, $S_2(r) = \langle \left| \delta u (r) \right|^2  \rangle$, plotted in Fig.~\ref{fig:exp_2_180} for $\Omega = 1.13$~rad~s$^{-1}$. This exponent is related to the distribution of energy among scales: larger values of $\zeta_2$ indicate a favored energy distribution toward larger scales. Significant oscillations of $\zeta_2$ are present, with a period equal to the tank rotation period, indicating the presence of inertial modes. Those inertial modes have been previously detected from oscillations in the decay of the kinetic energy by Morize et al.,\cite{Morize06} and their temporal spectrum has been analyzed in details by Bewley et al.\cite{Bewley07} Since we are interested here in the turbulent fluctuations that superimpose to those slow modes, we have computed the turbulent velocity fields $\tilde{\bf u}^\alpha({\bf x},t) = {\bf u}^\alpha({\bf x},t) - \langle {\bf u}^\alpha({\bf x},t) \rangle_\alpha$, where $\alpha$ denotes the realization and $\langle \cdot \rangle_\alpha$ is the ensemble average over the whole data set at a given time $t$ after the grid translation. The time evolution of the corrected exponent $\tilde{\zeta}_2$, measured from the scaling of the turbulent component of the SF, $\tilde{S}_2(r) = \langle \left| \delta \tilde{u} (r) \right|^2  \rangle$ (also plotted in Fig.~\ref{fig:exp_2_180}), is found to follow approximately the lower bound of the oscillations of the raw exponent $\zeta_2$.  One may conclude that the inertial mode, by superimposing a large scale modulation to the turbulence, leads to an increased raw exponent $\zeta_2$, of order of 10\%. In the following we will discard this slow inertial component of the flow and we will focus on the scaling of the turbulent flow component.

The corrected exponent, hereafter simply noted $\zeta_2$, is found to gradually increase during the decay, starting from values close to 2/3 at $t \simeq t_0$, as expected for an initial state weakly affected by rotation, and increasing up to $1.4 \pm 0.05$ at $t \simeq t_c$, reflecting the growing importance of the large scales compared to the small ones. This behavior compares well with the gradual steepening of the energy spectrum  reported by Morize et al.,\cite{Morize05} with a spectral exponent $p$ increasing from 1.7 to $2.3 \pm 0.1$ during the decay [dimensional analysis gives $\zeta_2 = p-1$, with $E(k) \sim k^{-p}$ the one-dimensional spectrum computed from the horizontal velocity and $k$ the horizontal wavenumber]. Beyond $t_c$,  the energy decreases exponentially as the result of the dissipation by the inertial waves, and no scaling range could be defined from the power spectrum.\cite{Morize05,Morize06} In the following we restrict to times $t < t_c$, where a correct scaling over an appreciable range of scales is observed from both $S_2(r)$ and $E(k)$.

\begin{figure}[t]
\centerline{\includegraphics[width=7.5cm]{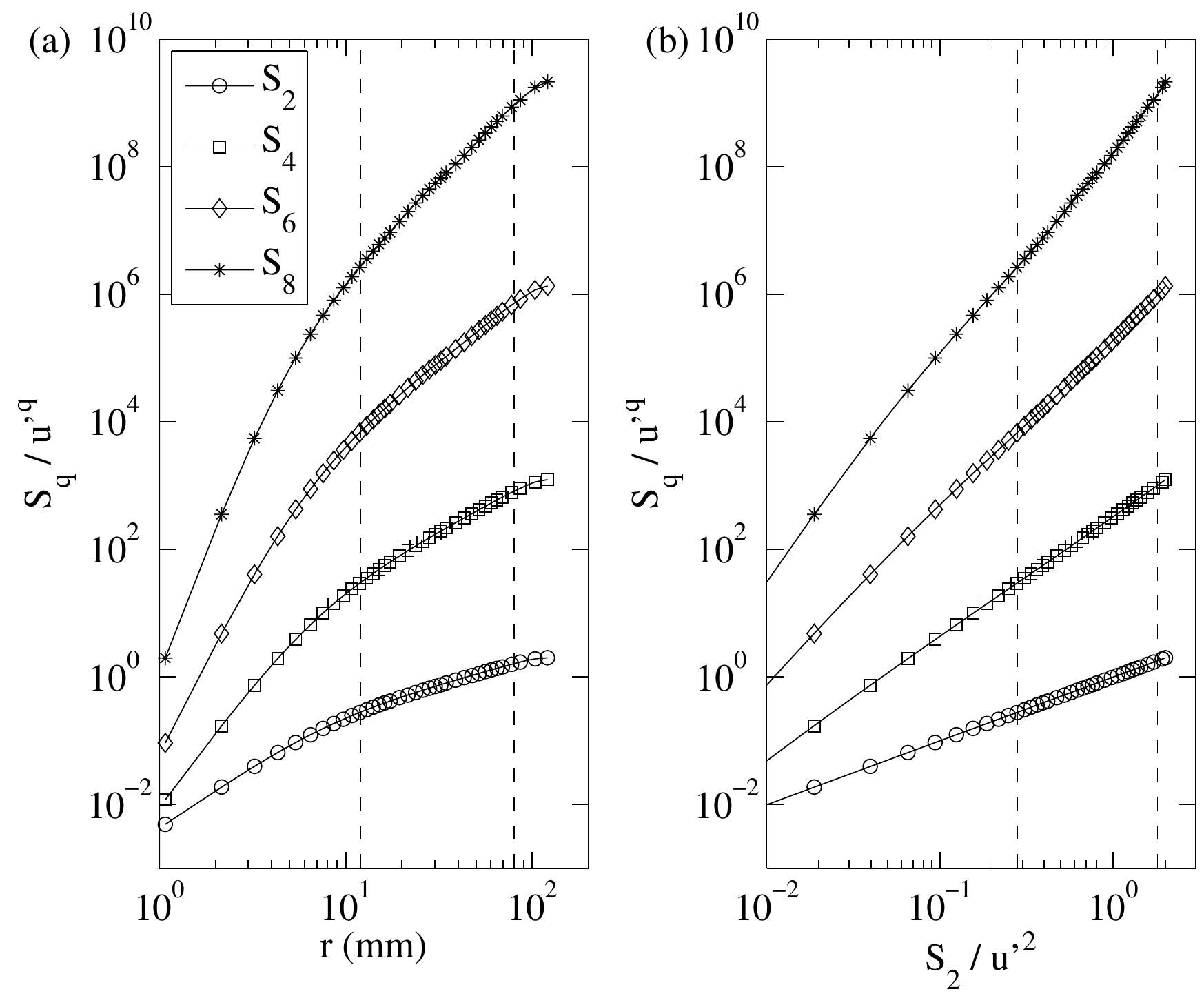}}
\caption{\footnotesize Structure functions for increasing orders, for $\Omega = 1.13$~rad~s$^{-1}$ and $\Omega t / 2 \pi = 2.7$, normalized by the velocity rms $u'(t)$ (a) plotted as a function of the separation $r$; (b) plotted as a function of $S_2$ (ESS method). The curves for $q=4,6,8$ have been vertically shifted by factors $10^2, 10^4$ and $10^6$ for visibility. The dashed lines show the range where the exponents are fitted.}
\label{fig:sq}
\end{figure}

We now turn to the higher order SFs.  Figure~\ref{fig:sq}(a), where SFs up to order $q=8$ are plotted at a given time $t$, show power laws for intermediate scales, here for 12~mm~$<r<80$~mm. It is worth pointing that the determination of the highest measurable order and its uncertainty for a given sample size is a delicate issue. The highest order for converged SF is determined by visual inspection of the truncated integral
\begin{equation}
C_q(r;\delta u^*) = \int_{-\delta u^*}^{\delta u^*}  p(\delta u) \left| \delta u (r)\right|  ^q \, d \delta u,
\label{eq:cq}
\end{equation}
which increases up to $S_q(r)$ as the cutoff $\delta u^*$ is increased.\cite{fn_integrand} Here $p$ is the probability density function (pdf) of the velocity increment $\delta u$.  For large separations and/or moderate orders, $C_q$ increases smoothly towards a well defined plateau as $\delta u^* \rightarrow \infty$, indicating a correct convergence of the SF.  On the other hand, smaller separations, $r < 10$~mm, show strong jumps when large velocity increments enter into the integral (\ref{eq:cq}). Those jumps may be due to either spurious vectors or insufficent statistics, and are the signature of an unconverged SF. According to this criterion, the range of separations $r$ ensuring a correct convergence of $S_q$ for $q>8$ is found too small for a reliable measurement of the scaling exponents, and measurements are restricted to order $q=8$. For orders $q \leq 8$, scales $r > 10$~mm were always correctly converged, allowing to safely define scaling exponents in the inertial range. Finally, the uncertainty $\Delta S_q(r)$ is estimated by plotting $S_q(r)$ at a given order and a given separation as a function of the sample size. Defining $\Delta S_q(r)$ as the standard deviation of $S_q(r)$ computed over the last third of the whole sample yields a relative error $\Delta S_q(r)/S_q(r)$ of 4\% for $q=4$ and 10\% for $q=8$, which is smaller than the symbol size in Fig.~\ref{fig:sq}.

\begin{figure}[t]
\centerline{\includegraphics[width=7cm]{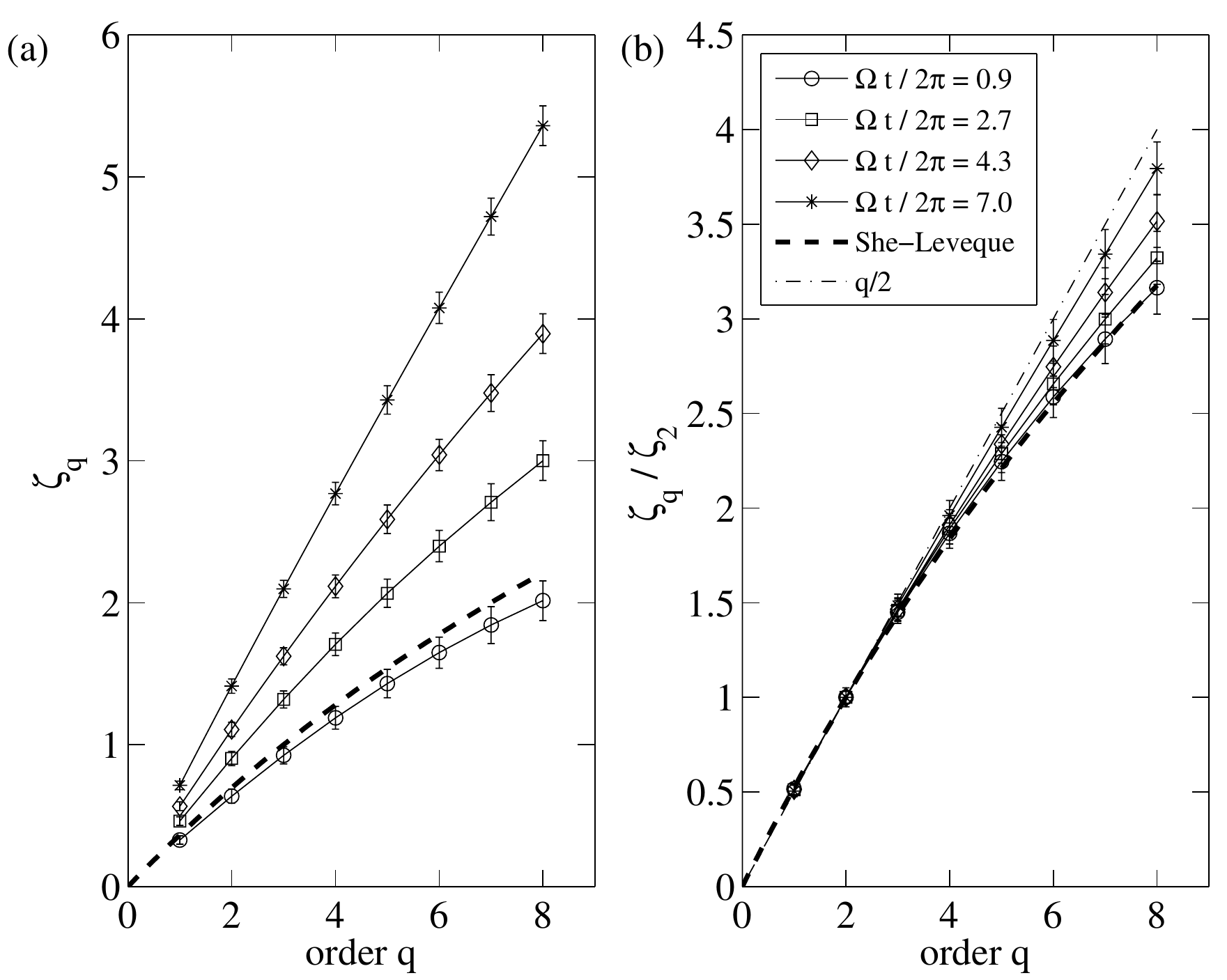}}
\caption{\footnotesize (a) Raw exponents $\zeta_q$, and (b) normalized exponents $\zeta_q / \zeta_2$ measured using ESS, at various times during the decay, for  $\Omega = 1.13$~rad~s$^{-1}$.}
\label{fig:exp}
\end{figure}

Figure~\ref{fig:exp} shows both the raw exponents $\zeta_q$ and the normalized exponents $\zeta_q / \zeta_2$ at different times during the decay. Those raw (resp. normalized) exponents are obtained from a linear least-square fit of $\log S_q$ versus $\log r$ (resp. $\log S_2$), following the Extended Self Similarity\cite{Benzi93} (ESS) procedure [see Fig.~\ref{fig:sq}(b)].\cite{fn_s2s3} The main contribution of the error bars for $\zeta_q$ is due to the uncertainty on the determination of the SF discussed above, $\Delta \zeta_q \simeq 2 (\Delta S_q / S_q) / \ln (r_2/r_1)$, where $r_1$ and $r_2$ are the lower and upper cutoffs of the scaling range, yielding $\Delta \zeta_4 \simeq 0.05$ and $\Delta \zeta_8 \simeq 0.14$. At the beginning of the decay, the effect of rotation is small and the exponents are indeed found very close to classical values for 3D non rotating turbulence.\cite{Frisch95} For comparison, the She-L\'ev\^eque\cite{She93} formula is also plotted, showing good agreement up to order $q=8$, giving confidence on the reliability of our PIV measurements. At larger times, the normalized exponents increase and become closer to the linear law $\zeta_q/\zeta_2 = q/2$, confirming the intermittency reduction induced by the background rotation. It is worth noting that the instantaneous Reynolds number at $t \simeq t_c$, $Re_M \simeq 360$ (see Table~\ref{tab:fp}), together with the correct scaling of the SF at that time, ensure that this intermittency reduction is not associated to the trivial scaling $\zeta_q = q$ (and hence $\zeta_q / \zeta_2 = q/2$) of a smooth velocity field.

The exponents at the end of the decay are comparable or even slightly larger than those reported by M\"uller and Thiele,\cite{Muller07} although their macro-Rossby numbers (0.01 and 0.05) are slightly lower and their Reynolds number (2300 and 4000) significantly larger than the present ones (note that the non-dimensional numbers here are based on the mesh size $M$, which underestimates the true integral scale). It must also be noted that the present exponents differ from the strictly linear law $q/2$ reported by Baroud \etal\cite{Baroud02} for similar Rossby numbers.  This slight discrepancy may be due to the different forcing mechanisms: In the present experiment, the initial turbulence produced by the grid translation is approximately isotropic, and rotation gradually breaks this initial isotropy in the course of the decay. In the experiment by Baroud \etal,\cite{Baroud02} turbulence is maintained by radial jets originating from a circular array of holes, generating a strong radial flow deflected by the Coriolis force. This forcing scheme is likely to produce an anistropic, partially two-dimensional flow, even in the absence of rotation. When rotation is present, this forcing probably reinforces the two-dimensional character of the turbulence, resulting in strictly non-intermittent exponents.

\begin{figure}[t]
\centerline{\includegraphics[width=6cm]{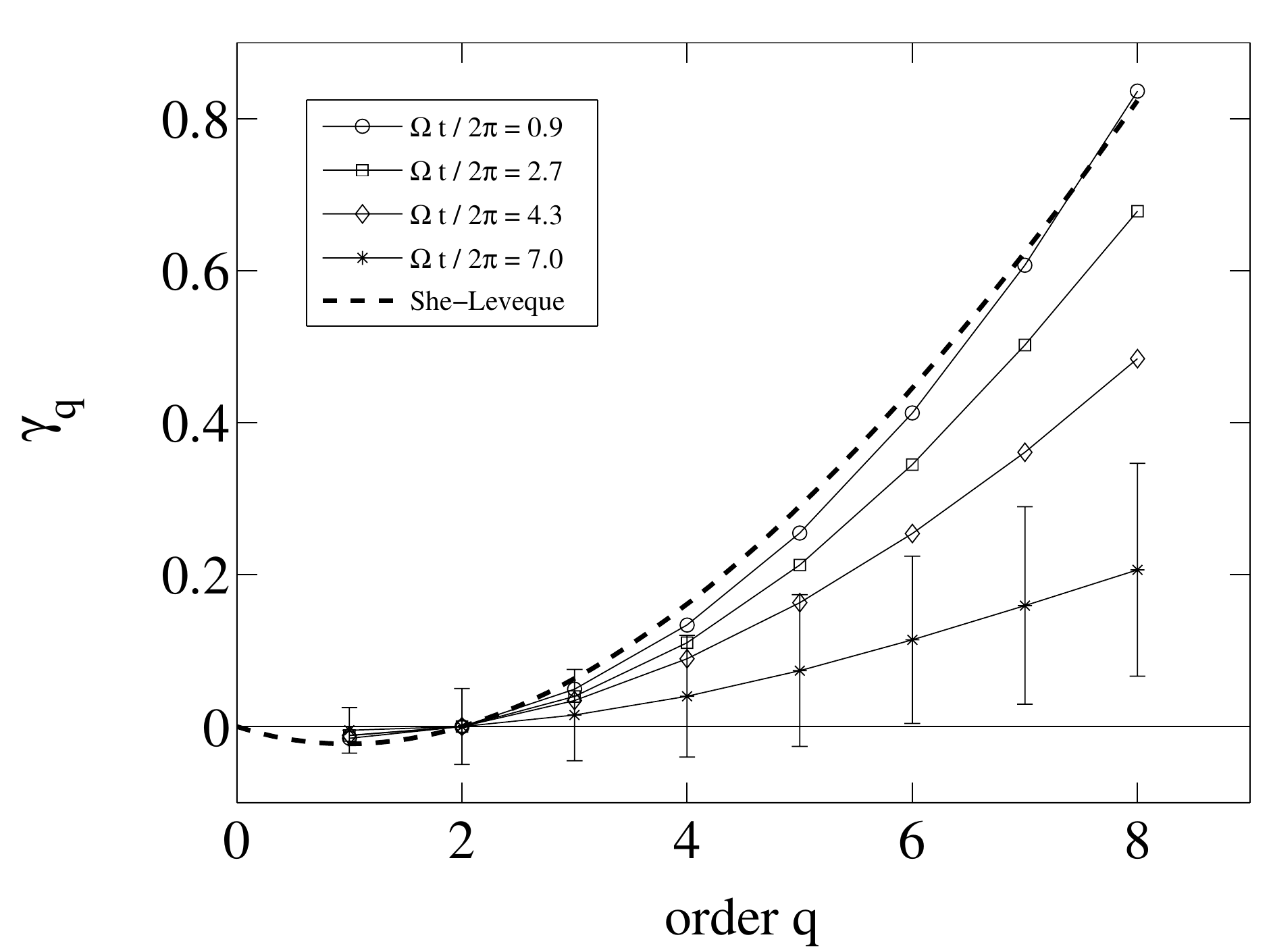}}
\caption{\footnotesize Intermittency factors $\gamma_q = q/2 - \zeta_q / \zeta_2$ (same data as in Fig.~\ref{fig:exp}). Error bars are only shown for the last curve ($*$) for clarity.}
\label{fig:if}
\end{figure}

\begin{figure}[t]
\centerline{\includegraphics[width=7cm]{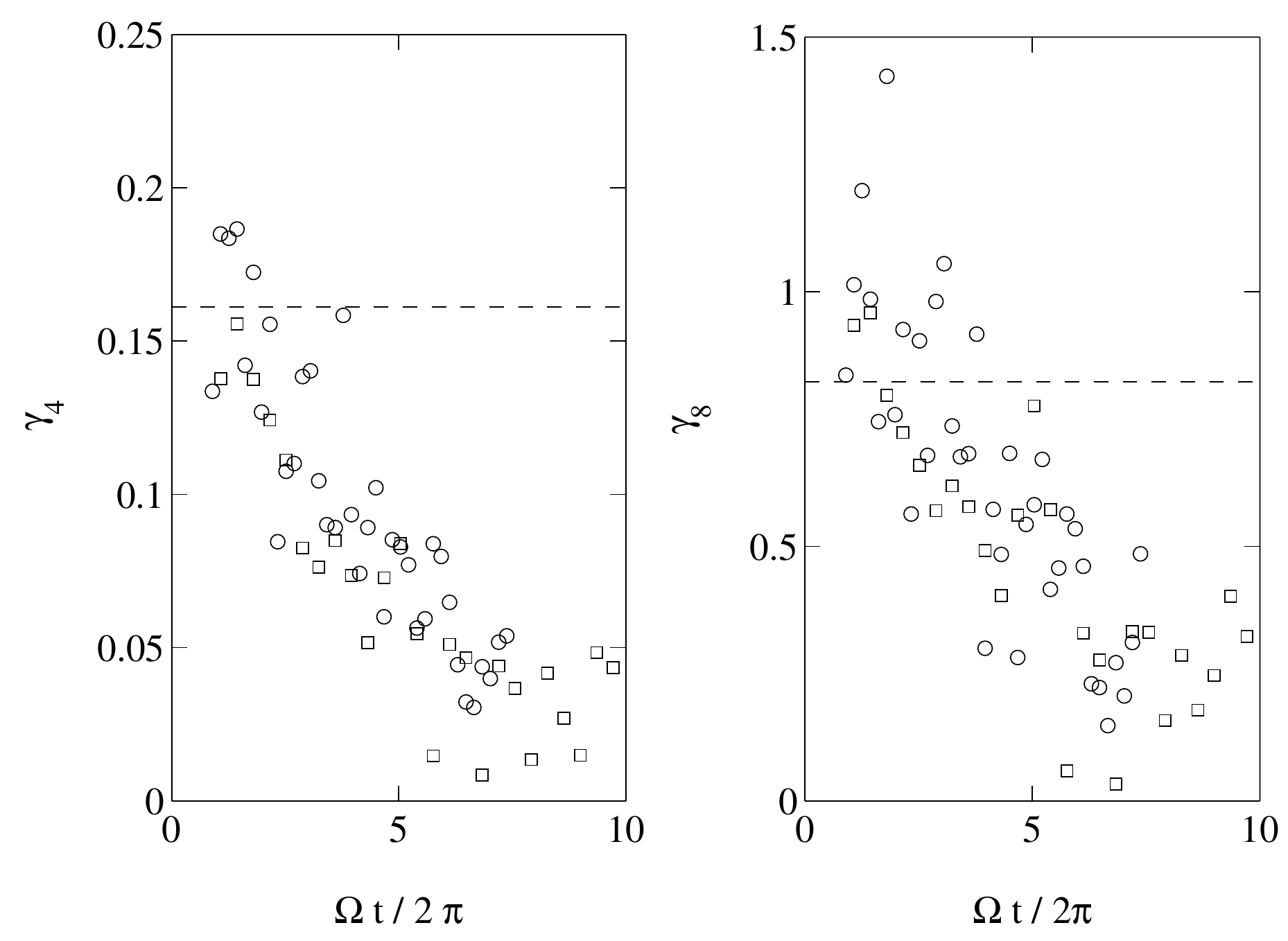}}
\caption{\footnotesize Time evolution of the 4th and 8th-order intermittency factors for the two experiments. $\circ$, $\Omega = 1.13$~rad~s$^{-1}$; $\square$, $\Omega = 2.26$~rad~s$^{-1}$. The dashed lines show the intermittency factors from the She-L\'ev\`eque model, $\gamma_4 = 0.161$ and $\gamma_8 = 0.824$. The error bars (not shown) are of the order of the scatter, $\Delta \gamma_4 = 0.05$ and $\Delta \gamma_8 = 0.14$.}
\label{fig:gamma}
\end{figure}

The reduction of intermittency during the decay is best appreciated from the intermittency factors $\gamma_q = q/2 - \zeta_q / \zeta_2$ (Figs.~\ref{fig:if} and \ref{fig:gamma}), which vanish for non-intermittent fluctuations.  Although the scatter is important on these quantities (of the order of $\Delta \zeta_q$), a clear trend towards smaller intermittency is present. It is interesting to note the approximate collapse of the data from the two rotation rates, suggesting that $\Omega^{-1}$ is the relevant time scale for the intermittency reduction.

The fact that the factors $\gamma_q$ start decreasing from the beginning of the decay is probably due to the low instantaneous Rossby numbers when $t \simeq t_0$ (see Table~\ref{tab:fp}). A crossover between constant $\gamma_q$ at early time and a decrease at larger times would be actually expected for larger grid Rossby number $Ro_g$. However, a large grid Reynolds number $Re_g$ is required for a developed turbulence to remain throughout the self-similar decay regime, up to the Ekman cutoff $t \simeq t_c$, limiting the maximum initial $Ro_g$ at fixed rotation rate and grid size. We finally note that, extrapolating the trend towards $\gamma_q \rightarrow 0$ in Fig.~\ref{fig:gamma}, suggests that the upper bound $t_c$ of the self-similar decay regime, in our experimental conditions, prevents from a clear observation of a vanishing intermittency, which may occur after 10-15 tank rotations.

To summarize, our measurements of SF in decaying rotating turbulence show a strong increase of the exponents $\zeta_q$ during the decay, which essentially follows the increase of the second-order exponent $\zeta_2$.  It is worth noting that values for $\zeta_2$ larger than 1 are found, in contradiction with the $S_2(r) \sim r$ [i.e., $E(k) \sim k^{-2}$] phenomenological law for rotating turbulence, derived under the assumption of nonlinear interactions governed by the timescale $\Omega^{-1}$.\cite{Zhou95,Muller07} Once normalized by $\zeta_2$, a marked increase of $\zeta_q / \zeta_2$ is observed,  a clear signature of a reduction of intermittency induced by the background rotation. This intermittency reduction is comparable to the one reported in a forced DNS by M\"uller and Thiele,\cite{Muller07} but it is less pronounced than in the forced experiment by Baroud \etal,\cite{Baroud02} probably because of the limited temporal range of self-similar decay due to the Ekman dissipation regime, which is specific to the decaying case.

We acknowledge L. Chevillard, S. Galtier, W.C. M\"uller, M. Rabaud and J. Rupper-Felsot for fruitful discussions.  This work was supported by the ANR grant no. 06-BLAN-0363-01 ``HiSpeedPIV''.


\begin{thebibliography}{99}
\bibitem{fn_integrand}
Instead of the convergence test based on Eq.~(\ref{eq:cq}), a similar test based on the integrand itself, $p(\delta u) \left| \delta u (r)\right|^q$ as $\delta u \rightarrow \infty$, has also been used in the literature. However, the visual quality of convergence from this quantity strongly depends on the choice of the bin width used to compute the pdf, especially for the far tails where empty bins are present, so criterion based on the truncated integral (\ref{eq:cq}) were found more reliable.
\bibitem{fn_s2s3}
For non-rotating turbulence, the ESS procedures is based upon $S_3$ instead of $S_2$, because of the prediction $\zeta_3=1$ from the Kolmogorov's 4/5 law. No such result applies for rotating turbulence, and $S_2$ were found more reliable because sign changes are present in $S_3(r)$ (computed without absolute values), associated to energy flux reversals.\cite{Morize05}
\end{thebibliography}
\end{document}